\documentclass[
unsortedaddress,
twocolumn,
amsmath,
amssymb,
aps,
prl,
]{revtex4-1}
\usepackage{braket}
\usepackage{graphicx}
\usepackage{lipsum}
\usepackage{dcolumn}
\usepackage{bm,color}
\usepackage[normalem]{ulem}

\usepackage{amsfonts}

\usepackage{epigraph}

\usepackage{etoolbox}
\makeatletter
\newlength\epitextskip
\pretocmd{\@epitext}{\em}{}{}
\apptocmd{\@epitext}{\em}{}{}
\patchcmd{\epigraph}{\@epitext{#1}\\}{\@epitext{#1}\\[\epitextskip]}{}{}
\makeatother

\begin{document}


\title{Non-Hermitian Realization of Quantum Dynamics on Embedded Manifolds}

\author{Samuel Alperin}

\affiliation{\vspace{1.25mm} \mbox{Los Alamos National Laboratory, Los Alamos, New Mexico 87545, USA}}

\begin{abstract}
We show that the Floquet Hamiltonian of a quantum particle driven by a general time-periodic imaginary potential is exactly equivalent, at stroboscopic times, to the Hamiltonian of a free particle constrained to a curved Riemannian manifold with fixed embedding. We illustrate the construction for a sinusoidal drive and for the torus of revolution, and outline how the framework can guide experimental design of curved-space quantum dynamics. Our results unify non-Hermitian Floquet physics with spectral geometry and provide a general recipe for engineering quantum dynamics on embedded manifolds.
\end{abstract}

\maketitle

The question of quantum mechanics on a manifold has been one of immense theoretical interest since the earliest days of quantum theory. Schrodinger's original approach was to simply replace the Laplacian $\nabla^2$ in the kinetic energy operator with its curved-space generalization, the Laplace-Beltrami operator $\widetilde{\nabla}^2_g$, defined on Riemannian manifold with metric $g$. In this approach, the effect of the geometry of the background space is entirely characterized by the metric, and thus by the local, or \textit{intrinsic}, geometry of the manifold.
However, while intrinsic geometric constraints are sufficient to define classical theories over arbitrary Riemannian manifolds, quantum mechanics is not a local theory, and  without further constraints, the quantization of classical theories over manifolds suffers from ambiguities \cite{JensenKoppe1971,daCosta1981,Ortix2015,FerrariCuoghi2008,JensenKoppe1971,daCosta1981}. Another approach is one in which a quantum theory defined in $\mathbb{R}^{m}$ is constrained by an infinitely strong potential normal to the surface of an embedded submanifold, $\mathcal{M}^n$. The explicit embedding of $\mathcal{M}^n$ fixes not only its intrinsic geometry, but also its extrinsic geometry. The difference can be understood by considering distinct embeddings of isometric manifolds, for example the Euclidean plane and the cylinder represent different embeddings with the same metric, having distinct extrinsic geometry  but equivalent intrinsic geometry. Such an approach, developed by Jensen and Koppe and independently by da Costa, successfully avoids all of the ambiguities \cite{JensenKoppe1971,daCosta1981,Ortix2015,FerrariCuoghi2008} associated with Schrodinger's local-geometric approach, and has led to numerous generalizations \cite{Encinosa1998,FerrariCuoghi2008,Ortix2015,Cantele2000,Schuster2003,Taira2010}. Such work has led to deep and surprising results, such as the discovery of geometrically emergent gauge fields in the theories of scalar particles constrained to manifolds embedded in higher dimensions \cite{Schuster2003}. However, even in low dimensions the effects of both intrinsic and extrinsic geometry are very much physical, and with recent developments in the fabrication of physical systems with nontrivial geometries, the importance of understanding the physics of quantum theories on embedded submanifolds has increased dramatically, being experimentally relevant in in carbon nanotubes and nanoribbons \cite{Shima1999,Onoe2012}, semiconductor heterostructures \cite{FerrariCuoghi2008,Ortix2015}, photonic lattices \cite{Kartashov2021,Schwartz2007}, topological surface states \cite{Ortix2010,EntinMagarill2001}, and cold atoms \cite{Amico2005,Boada2011}.


Meanwhile, over the past quarter century there has been a dramatic flurry of interest in both the theory and experimental realization of quantum mechanical systems with non-Hermitian Hamiltonians. As was first shown in the seminal work of Bender and Boetcher, there exists a large class of non-Hermitian Hamiltonians which have strictly real spectra \cite{BenderBoettcher1998,Graefe2012,ElGanainy2018} \cite{BenderBoettcher1998}. The realization of these Hamiltonians has allowed for the rapid discovery of a wide range of striking phenomena, from one-way invisibility to topological lasing \cite{RudnerLevitov2009,Harari2018,Hodaei2014,RudnerLevitov2009}. Originally it appeared that all non-Hermitian Hamiltonian operators with real spectra  had the property of parity-time ($\mathcal{PT}$) symmetry a simple and geometrically interpretable property \cite{BenderBoettcher1998,Graefe2012,ElGanainy2018}. However, it was soon proven that PT symmetry is neither necessary or sufficient for a Hamiltonian to have real spectra, with the actual necessary and sufficient condition being $\eta$-pseudo-Hermitian symmetry for some metric $\eta$ \cite{Mostafazadeh2010}.
 
Recently, a relationship between non-Hermiticity and geometry has been suggested by the realization that the continuum limits of specific non-Hermitian tight-binding lattice models can inherit hyperbolic structure  within the effective kinetic operators \cite{lv2022curving,wang2022duality}. In this Letter, unify and extend these earlier examples by providing a general Floquet similarity that applies to arbitrary time-periodic imaginary drives and yields an exact operator-level duality with quantum mechanics on manifolds embedded in generally higher dimensional spaces.   Finally, we construct a simple methodology by which an imaginary Floquet drive can be engineered such that the curved-space dual has a prescribed geometry, and construct explicit examples.






\noindent\textit{General Duality} -- 
We begin with the equation of motion for a quantum particle in the ordinary flat space $\mathbb{R}^m$, subject to a generally complex, time-periodic (Floquet) potential $\Gamma$ in addition to time-invariant potential $V$, which takes the form

\begin{equation}
   i \partial_t \psi=-\nabla^2 \psi + \Gamma(\mathbf{r},t)\psi-V(\mathbf{r})\psi,
    \label{physical}
\end{equation}
with wavefunction $\psi$, where $\mathbf{r}$ is the position vector,  $\nabla^2$ is the ordinary Laplacian operator, and where we have set $\hbar=2m=1$.
Assuming the spatial and temporal structure of the Floquet drive to be separable, we can rewrite it as $\Gamma(\mathbf{r},t)=\partial_t {\gamma}(\mathbf{r},t)=\bar\gamma(\mathbf{r})\dot{f}(t)$, where $ f(t)= f(t+T)$ is some time-periodic function with period $T$. Moving into the frame of the periodic micromotion with the transformation$\psi=P(t)\varphi$ where we define
\begin{equation}
P(t)=\exp\!\big[f(t)\,\bar\gamma(\mathbf r)\big],\qquad P(t+T)=P(t),
\end{equation}
we have that $\varphi$ obeys $i\partial_t\varphi=H'(t)\varphi$ with
\begin{align}
H'(t)&=P^{-1}(t)\big(H(t)-i\partial_t\big)P(t)\nonumber\\
&= -\nabla^2
-2f(t)\,\nabla\bar\gamma\!\cdot\!\nabla
- f(t)\,\nabla^2\bar\gamma \nonumber\\
&\quad - \big(f(t)\nabla\bar\gamma\big)^2
- V(\mathbf r).
\label{eq:Hprime}
\end{align}
For notational simplicity this is rewritten as
\begin{equation}
H'(t)=A+f(t)\,B+f(t)^2\,C,
\label{eq:ABC}
\end{equation}
where we define
\begin{align}\label{abc}
A&=-\nabla^2-V, \nonumber\\
B&=-2\,\nabla\bar\gamma\!\cdot\!\nabla-\nabla^2\bar\gamma, \nonumber\\
C&=-(\nabla\bar\gamma)^2.
\end{align}

The exact stroboscopic evolution over one period T is given by
\begin{equation}
U'(T)=\mathcal T\exp\!\Big(-i\!\int_0^T\!H'(t)\,dt\Big)\equiv e^{-iH_F T},
\label{eq:monodromy}
\end{equation}
which fully defines $H_F$ up to the natural degeneracy in quasienergy.

We now evaluate $H_F$ via the Magnus expansion \cite{Blanes2009}:
\begin{align}
H_F &= \tfrac{1}{T}\,\Omega_1+\tfrac{1}{T}\,\Omega_2+\tfrac{1}{T}\,\Omega_3+\cdots,
\label{eq:Magnus-series}\\
\Omega_1&=\int_0^T\!H'(t_1)\,dt_1, \nonumber\\
\Omega_2&=\tfrac12\int_0^T\!dt_1\!\int_0^{t_1}\![H'(t_1),H'(t_2)]\,dt_2. \nonumber
\end{align}
Defining the temporal moments
\begin{equation}
m_1=\frac{1}{T}\!\int_0^T\!f(t)\,dt,\qquad
m_2=\frac{1}{T}\!\int_0^T\!f(t)^2\,dt,
\end{equation}
the leading contribution to the Hamiltonian is
\begin{equation}
H_F^{(1)}=A+m_1 B+m_2 C.
\label{eq:HF1}
\end{equation}

For the second order Magnus term, we note that $[B,B]=[C,C]=0$, so that the algebra can be found to close over $[A,B]$, $[A,C]$, $[B,C]$, so that
\begin{equation}
\frac{\Omega_2}{T}=\alpha_{AB}[A,B]+\alpha_{AC}[A,C]+\alpha_{BC}[B,C].
\label{eq:HF2-structure}
\end{equation}
With  coefficients which can be written as exact functionals of $f$:
\begin{align}
\alpha_{AB}&=\frac{1}{T}\!\int_0^T\!(t-\tfrac{T}{2})\,f(t)\,dt, \label{eq:alphaAB}\\
\alpha_{AC}&=\frac{1}{T}\!\int_0^T\!(t-\tfrac{T}{2})\,f(t)^2\,dt, \label{eq:alphaAC}\\
\alpha_{BC}&=\frac{1}{2T}
\begin{aligned}[t]
  &\int_0^T\!\!\int_0^T 
  \frac{\mathrm{sgn}(t_1-t_2)}{2}\, f(t_1)f(t_2) \\
  &\times \big(f(t_2)-f(t_1)\big)\, dt_1\,dt_2
\end{aligned}
\label{eq:alphaBC}
\end{align}
Here, we note that the $\alpha$-terms vanish exactly for time-symmetric drives, and vanish approximately for any drive in the high-frequency limit. As a result, we assume symmetric driving and discard the higher-order Magnus terms, leaving us with the exact stroboscopic Hamiltonian

Substituting the terms defined in Eqs. \ref{abc} and absorbing all differential operators into one Laplace-Beltrami operator exactly as in the main text, the stroboscopic Hamiltonian is given, to second order, by
\begin{align}
i\partial_t\varphi
&= -\nabla_g^2\,\varphi + (K+M^2)\varphi \nonumber\\
&\quad +\alpha_{AB}[A,B]\varphi
 +\alpha_{AC}[A,C]\varphi
 +\alpha_{BC}[B,C]\varphi.
\label{eq:central}
\end{align}

For reasons that will soon become clear, we define the coordinate transformation $\mathfrak{M}:\mathbf{r}(x_0,x_1,...)\rightarrow \mathbf{z}(x'_0,x'_1,...)$ such that the spatial differential operator in Eq. \ref{tavg} takes the form

\begin{equation}
    \nabla^2_r+2\bar f \nabla_r \gamma\cdot \nabla_r = \kappa(z) \nabla_z^2,
    \label{conseq}
\end{equation}
where $\nabla_r=\sum_i\partial_{x_i}$ and  $\nabla_z=\sum_i\partial_{x'_i}$ are the Euclidean Laplacians in coordinates $\mathbf{r}$ and $\mathbf{z}$, respectively.


It is easy to show that the transformation $\mathfrak{M}$ is determined by the differential consistency condition

\begin{equation}
    \nabla_r \mathbf{z} \nabla_r\kappa(\mathbf{z})=2\kappa \nabla_r \gamma,
    \label{consist2}
\end{equation}
which is satisfied only for 

\begin{equation}
    \kappa(\mathbf{x})=\left(\nabla_\mathbf{x}\mathbf{z}(\mathbf{x})\right)^2
\end{equation}
and 
\begin{equation}
    \mathbf{z}(\mathbf{x})=\int_1^\mathbf{x}\rm{e}^{2\gamma(\mathbf{x}')d\mathbf{x}'}.
\end{equation}
Therefore given a particular form of $\gamma(\mathbf{x})$, then as long as the map $\mathfrak{M}$ is invertible, there exists some fixed function $\kappa(x)$ and map $\mathfrak{M}^{-1}:\mathbf{z}\rightarrow \mathbf{x}$ such that we can rewrite Eq. \ref{tavg} as

\begin{equation}
   i  \partial_t \varphi=-\kappa(z) \nabla_z^2 \varphi
  +\left(\bar f\nabla_x^2 \gamma +(\bar f\nabla_x\gamma)^2\right)\varphi+V\varphi.
  \label{curvgen}
\end{equation}

From here we note that the kinetic energy operator in Eq. \ref{curvgen}, $\kappa(z)\nabla_z^2$, is identical to the Laplace-Beltrami operator $\nabla_g^2$ for conformally flat metric $g$ defined such that 

\begin{equation}
     ds^2=\frac{1}{\kappa}\left(dx_0^{'2}+dx_1^{'2}+...\right)=\frac{1}{\kappa}\mathbf{\eta}.
\end{equation}

Rewriting the spatial  derivatives of $\gamma$ within the effective potential in terms of the new coordinates $\mathbf{z}$, we find that

\begin{equation}
   i  \partial_t \varphi=-\nabla_g^2  \varphi
  +\left(\rm{e}^{4\gamma}\bar f\nabla_z^2 \gamma +\rm{e}^{8\gamma}(\bar f\nabla_z\gamma)^2\right)\varphi+V\varphi.
  \label{final}
\end{equation}
However, noticing that the Gaussian, or intrinsic, curvature of the manifold characterized by metric $g$ takes the exact form 

\begin{equation}
    K=\rm{e}^{4\gamma}\bar f\nabla_z^2 \gamma +\rm{e}^{8\gamma}(\bar f\nabla_z\gamma)^2,
\end{equation}
we find that Eq. \ref{final} is exactly equivalent to the general form of the equation of motion of a quantum particle constrained an embedded manifold,
\begin{equation}
    i \partial_t \varphi=-\nabla_g^2  \varphi
  +\left(M^2+K\right)\varphi,
  \label{jaffe}
\end{equation}
where the square of the mean (intrinsic) curvature is fixed as $M^2=V$.
We therefore arrive at the central result of this Letter: the effective theory of particle in flat space under forcing $\Gamma$ (Eq. \ref{final}), is equivalent to the theory of a free particle constrained to a particular manifold, whose intrinsic geometry is set by the forcing of the underlying theory, and whose extrinsic geometry is fixed by the time invariant potential of that underlying theory. Thus, to engineer a physical system which is dual to a free theory on a particular embedded manifold, one follows the following general process: one solves Eq. \ref{consist2} by either fixing the form of the Floquet drive or by fixing the desired conformal curvature factor $\kappa$. Either way, this defines a coordinate transform $\mathfrak{M}$ which, if invertible, allows a map from the Floquet theory to its dual theory of a free particle on a manifold. We note that we may now also interpret our coordinate transforms geometrically with $x(z)$ being the \emph{optical length} of the metric $dz/\sqrt{\kappa_{\rm tar}(z)}$, and $\bar\gamma$ representing the logarithm of the conformal factor scaled by $1/(4\bar f)$. Further, we note that the structure of the embedding of the manifold is determined entirely by $V$: setting $V=0$ necessarily yields a theory on a minimal surface.

\noindent\textit{Geometry of Spectral Reality} -- 
One of the central puzzles in non-Hermitian quantum mechanics is why certain classes of Hamiltonians, most famously those with $\mathcal PT$ symmetry, can exhibit entirely real spectra despite their non-Hermitian form. In our construction this phenomenon admits a simple and transparent explanation. By design, the driven flat–space Hamiltonian is mapped, via the periodic similarity transform $P(t)$ and the conformal change of variables $x\mapsto z$, onto a dual operator
\begin{equation}
H_{\rm geom}=-\nabla_g^2+K(z)+M^2(z),
\label{eq:Hgeom-spectral}
\end{equation}
which acts on wavefunctions in the Hilbert space $L^2(\sqrt{g}\,dz)$ of the curved metric $ds^2=\kappa(z)^{-1}dz^2$. Crucially, $H_{\rm geom}$ is manifestly self–adjoint in this space: the Laplace--Beltrami operator is Hermitian with respect to the curved measure, while the scalars $K(z)$ and $M^2(z)$ are real functions of position. Thus the stroboscopic quasienergies, which are the eigenvalues of $H_{\rm geom}$, are automatically real. 

From this perspective, the ``mystery'' of reality in the non-Hermitian frame is resolved by noting that the apparent non-Hermiticity is an artifact of using the wrong inner product. The micromotion similarity $S=P(t)\cdot\mathcal U$ (where $\mathcal U$ is the unitary implementing the $x\mapsto z$ change of measure) relates the physical stroboscopic Hamiltonian $H_F$ to its geometric dual:
\begin{equation}
H_{\rm geom}=S\,H_F\,S^{-1},\qquad 
\eta:=S^\dagger S>0.
\end{equation}
It follows that $H_F$ is $\eta$--pseudo–Hermitian, satisfying $H_F^\dagger\eta=\eta H_F$. The non-Hermitian drive thus produces a Floquet Hamiltonian whose spectrum is real because it is similar to a Hermitian operator, with the metric operator $\eta$ supplied by the geometry itself. 

In this light, $\mathcal PT$ symmetry is seen not as the fundamental reason for spectral reality, but as a particular \emph{manifestation} of the underlying Hermiticity of the curved–space dual. If the manifold admits an involutive isometry $\mathcal I:z\mapsto -z$ and the scalar data $K$ and $M^2$ are even under this reflection, then $H_{\rm geom}$ commutes with the antiunitary parity–time operator $(\mathsf P\mathsf T)\psi(z)=\psi(-z)^*$. In the flat–space picture this corresponds to the conventional $\mathcal PT$ symmetry of the driven Hamiltonian. When that isometry is present the spectrum is real and $\mathcal PT$ is unbroken; when it is absent or explicitly violated by the geometry, $\mathcal PT$ is broken and complex quasienergies appear. The transition has a clear geometric signature: it is the loss of a discrete isometry in the dual manifold.

The geometric interpretation therefore recasts the reality of quasienergy spectra as a straightforward consequence of ordinary Hermiticity in curved space, while providing a simple geometric diagnostic for $\mathcal PT$ transitions. What appears in the laboratory frame as a delicate balance of gain and loss is, in the dual frame, nothing more exotic than a free particle exploring a curved manifold with reflection symmetry.

In the broken phase this geometric picture remains equally transparent. Once the reflection isometry of the manifold is lost, either because the conformal factor $\kappa(z)$ or the embedding potential $V(z)$ develops an odd component, the Hamiltonian $H_{\rm geom}$ no longer commutes with the parity operator. Although $H_{\rm geom}$ itself remains Hermitian, the similarity back to the flat--space stroboscopic Hamiltonian $H_F$ ceases to preserve a real spectrum: the metric operator $\eta=S^\dagger S$ is no longer parity–invariant, and pairs of complex–conjugate quasienergies emerge. Geometrically, the eigenmodes respond to the broken symmetry by localizing asymmetrically in the curved coordinate $z$, piling up preferentially on one side of the manifold. This reproduces in the dual language the characteristic non-Hermitian skin effect and asymmetric transport seen in the laboratory frame. Thus the complexification of quasienergies in the broken phase acquires a clear geometric interpretation: it is the spectral fingerprint of curvature and embedding data that fail to admit a reflection symmetry, and the associated eigenfunctions visibly break parity in the curved space itself.

On manifolds with boundary, the same reasoning applies provided one imposes self–adjoint boundary conditions (Dirichlet, Neumann, or real Robin); non–self–adjoint terminations correspond to PT–broken phases in the laboratory frame.

\bigskip
\noindent\textit{Engineering Dual Manifolds --} We now turn to the problem of engineering the Floquet drive so as to generate dynamics equivalent to those of a free particle constrained to the surface of a prescribed Riemannian manifold.

We start with the stroboscopic differential operator, which at leading order takes the following form in the laboratory frame:
\[
\mathcal K_{(1)}=\partial_x^2+2\,\bar f\,\bar\gamma'(x)\,\partial_x.
\]
By construction,  there exists a strictly monotonic change of variables $x\mapsto z$ such that
\[
\mathcal K_{(1)}=\kappa(z)\,\partial_z^2,
\qquad 
\text{equivalently}\qquad 
ds^2=\frac{1}{\kappa(z)}\,dz^2,
\]
i.e.\ the kinetic term is the Laplace--Beltrami operator of a \emph{conformally flat} one--dimensional metric with conformal factor $\kappa(z)>0$. 
The change of variables is fixed by the identities
\begin{equation}
\frac{dz}{dx}=e^{\,2\bar f\,\bar\gamma(x)}
\qquad\Longleftrightarrow\qquad
\kappa=\Big(\frac{dz}{dx}\Big)^{\!2}=e^{\,4\bar f\,\bar\gamma}.
\label{eq:design-core}
\end{equation}
From \eqref{eq:design-core}, a fixed conformal factor $\kappa$ constrains the form of the drive profile $\bar\gamma$ as well as the coordinate map  $\mathcal{M}$. 
Suppose the metric of the dual manifold is prescribed as
\[
ds^2=\frac{1}{\kappa_{\rm tar}(z)}\,dz^2
\qquad(\text{with }\kappa_{\rm tar}(z)>0).
\]
Then the drive profile in the $z$--frame is 
\begin{equation}
\bar\gamma(z)=\frac{1}{4\bar f}\,\ln \kappa_{\rm tar}(z).
\label{eq:gamma-from-kappa-z}
\end{equation}
The form of the drive only makes physical sense in the laboratory frame, so we transform back to yield
\begin{equation}
\frac{dz}{dx}=e^{\,2\bar f\,\bar\gamma(z)}=\sqrt{\kappa_{\rm tar}(z)}
\quad\Longrightarrow\quad
x(z)=\int^{\,z}\!\frac{d\zeta}{\sqrt{\kappa_{\rm tar}(\zeta)}}.
\label{eq:x-of-z-from-kappa}
\end{equation}
Provided that $\kappa_{\rm tar}$ is smooth and strictly positive, $x(z)$ can be inverted to give $z(x)$. 
The Floquet drive in the laboratory frame  is then given by the composition
\begin{equation}
\quad \bar\gamma(x)\;=\;\frac{1}{4\bar f}\,\ln \kappa_{\rm tar}\!\big(z(x)\big),\qquad 
x(z)=\int^{\,z}\!\kappa_{\rm tar}(\zeta)^{-1/2}d\zeta.\quad
\label{eq:gamma-of-x-from-kappa-z}
\end{equation}

\bigskip
\noindent\textit{Example I -- Periodic Curvature:}
We first consider a prescribed, trivially embedded manifold  characterized by a sinusoidally modulated conformal factor, such that
\[
\kappa_{\rm tar}(z)=1+\lambda\cos(Qz)\qquad (|\lambda|<1),
\]
then \eqref{eq:gamma-from-kappa-z} gives
\[
\bar\gamma(z)=\frac{1}{4\bar f}\,\ln\!\big[1+\lambda\cos(Qz)\big],
\]
and \eqref{eq:x-of-z-from-kappa} yields
\[
x(z)=\int^{\,z}\!\frac{d\zeta}{\sqrt{1+\lambda\cos(Q\zeta)}}
\], which is an elementary elliptic integral.  
For weak modulation $|\lambda|\ll 1$, we can  expand as
\[
\bar\gamma(z)=\frac{\lambda}{4\bar f}\cos(Qz)+\mathcal O(\lambda^2),
\qquad
x(z)=z+\mathcal O(\lambda).
\]
Therefore, we have$\bar\gamma(x)=\frac{\lambda}{4\bar f}\cos(Qx)$ to leading order.
Interestingly, this is exactly the Hamiltonian found by taking the continuum limit of the off-diagonal Andre-Aubry-Harper model.

\medskip
\noindent\textit{Example II -- The Torus:} We now engineer a drive profile $\bar\gamma$ and a time-independent $V$ so that the stroboscopic Floquet Hamiltonian reproduces the dynamics of a particle constrained to the surface of a torus of revolution with major radius $R$ and minor radius $r$ ($R>r$). Write the usual embedding with angles $(\theta,\phi)\in(-\pi,\pi]\times(-\pi,\pi]$; in these coordinates the first fundamental form is \[ ds^2 = r^2\,d\theta^2 + \big(R+r\cos\theta\big)^2\,d\phi^2. \] To match our Floquet mapping, we first pass to \emph{isothermal} (conformal) coordinates so the metric is $ds^2=\Omega^2(u,v)\,(du^2+dv^2)$. For a surface of revolution $E(\theta)d\theta^2+G(\theta)d\phi^2$ with $E=r^2$, $G=(R+r\cos\theta)^2$, a standard choice is \[ v=\phi,\qquad u=u(\theta)\ \ \text{with}\ \ du=\frac{\sqrt{E}}{\sqrt{G}}\,d\theta=\frac{r}{R+r\cos\theta}\,d\theta. \] Integrating gives a closed form \begin{equation} u(\theta)=\frac{2r}{\sqrt{R^2-r^2}}\, \arctan\!\Bigg(\sqrt{\frac{R-r}{R+r}}\ \tan\frac{\theta}{2}\Bigg), \label{eq:u-of-theta} \end{equation} which is strictly monotonic in $\theta$ and periodic. In these coordinates the metric is conformal: \begin{equation} ds^2 = \Omega^2(u)\,\big(du^2+dv^2\big),\qquad \Omega^2(u)=\big(R+r\cos\theta(u)\big)^2. \label{eq:torus-conformal} \end{equation} Our Floquet construction realizes conformally flat metrics in the form $ds^2=\kappa^{-1}(\mathbf z)\,(dz_1^2+dz_2^2)$, with \[ \kappa(\mathbf z)=e^{4\bar f\,\bar\gamma(\mathbf z)}. \] Comparing with \eqref{eq:torus-conformal}, we target \[ \kappa_{\rm tar}(u,v)=\frac{1}{\Omega^2(u)}=\frac{1}{\big(R+r\cos\theta(u)\big)^2}. \] By the design rule $\bar\gamma=\tfrac{1}{4\bar f}\ln\kappa$, we therefore \emph{choose the drive in isothermal coordinates} as \begin{equation} \bar\gamma(u,v) = \frac{1}{4\bar f}\,\ln\!\Big[\kappa_{\rm tar}(u)\Big] = -\,\frac{1}{2\bar f}\,\ln\!\big(R+r\cos\theta(u)\big),  \label{eq:gamma-torus} \end{equation} with $u(\theta)$ given by \eqref{eq:u-of-theta}. With this choice, the leading Floquet kinetic operator becomes exactly the Laplace--Beltrami operator on the torus written in the $(u,v)$ chart: \[ -\kappa\,\nabla^2_{(u,v)}\ \equiv\ -\nabla_g^2,\qquad ds^2=\kappa^{-1}(u)\,(du^2+dv^2), \] i.e.\ the intrinsic metric is \emph{identical} to that of the geometric torus in conformal gauge.

It remains to choose the static potential $V$ to account for the extrinisc curvature that emerges in the quantum theory of the particle on the torus. The mean curvature of the torus $\mathbb{T}^2$ under standard embedding in $\mathbb{R}^3$ takes the form

\begin{equation}
    M=\frac{1}{2}\left(\frac{1}{r}+\frac{\cos\theta}{R+r\cos\theta}\right),
\end{equation}
so that by setting $\quad V(u,v)= -\,H\!\big(\theta(u)\big)^{\!2}\quad $ and transorming back into the laboratory frame, the total stroboscopic Hamiltonian is exactly the Hamiltonian for the free particle constrained to the naturally embedded torus.

The author acknowledges support from Los Alamos National Laboratory LDRD program grant 20230865PRD3.

\bibliographystyle{apsrev4-2}

\end{document}